\documentstyle[12pt,epsf]{article}
\newcommand{\beq}{\begin{equation}}
\newcommand{\eeq}{\end{equation}}
\newcommand{\beqa}{\begin{eqnarray}}
\newcommand{\eeqa}{\end{eqnarray}}
\newcommand{\boldtau}{\mbox{\boldmath $\tau$}}
\newcommand{\boldS}{\mbox{\boldmath $S$}}

\begin{document}

\begin{titlepage}


\hfill{DOE/ER/40561-57-INT99}

\hfill{TRI-PP-99-24}

\hfill{KRL MAP-248}

\hfill{NT@UW-99-30}

\vspace{1.0cm}

\begin{center}
{\Large {\bf Effective Theory of the Triton}}

\vspace{1.2cm}

{\large P.F. Bedaque$^{a,}$\footnote{{\tt bedaque@mocha.phys.washington.edu}},
H.-W. Hammer$^{b,}$\footnote{{\tt hammer@mps.ohio-state.edu}}$^,$\footnote{
Present address: Department of Physics, The Ohio State University,
Columbus, OH 43210, USA},
and U. van Kolck$^{c,d,}$\footnote{{\tt vankolck@krl.caltech.edu}}}

\vspace{1.2cm}
{\it
$^a$~Institute for Nuclear Theory,
 University of Washington, Seattle, WA 98195, USA
~\\$^b$TRIUMF, 4004 Wesbrook Mall,
Vancouver, B.C., Canada V6T 2A3
~\\$^c$ Kellogg Radiation Laboratory, 106-38,
 California Institute of Technology, \\
 Pasadena, CA 91125, USA\\
$^d$ Department of Physics,
 University of Washington, Seattle, WA 98195, USA}
\end{center}

\vspace{1cm}

\begin{abstract}
We apply the effective field theory approach
to the three-nucleon system. In particular, we consider 
$S=1/2$ neutron-deuteron scattering and the triton.
We show that in this channel a unique 
nonperturbative renormalization takes place which requires the 
introduction of a single three-body force at leading order.
With one fitted parameter we find a good description of low-energy data.
Invariance under the renormalization group
explains some universal features of the three-nucleon system
---such as the Thomas and Efimov effects and the Phillips line---
and the origin of $SU(4)$ symmetry in nuclei.
\end{abstract}

\vspace{2cm}
\vfill
\end{titlepage}

\section{Introduction}
Effective field theories (EFT's) are a powerful concept
designed to explore a separation of scales in physical 
systems \cite{gospel}. For example if the momenta $k$ of two
particles are much smaller than the inverse range of 
their interaction $1/R$, observables can be expanded in powers of $kR$.
Following the early work of Weinberg and others \cite{weinberg},
EFT's have become quite popular in nuclear physics \cite{nn98,monster}. 
Their application, however, is complicated
by the presence of shallow (quasi) bound states, which create 
a large scattering length $a\gg R$. In this fine-tuned
case, a simple perturbative expansion in $kR$ already
breaks down at rather small momenta $k\sim1/a$. In order to 
describe bound states with typical momenta $k\sim1/a$,
the range of the EFT has to be extended.
A certain class of diagrams has to be resummed which generates a 
new expansion in $kR$ where powers of $ka$ are kept to all orders.
For the two-nucleon system a power counting that incorporates
this resummation has been found recently \cite{1stooge,3musketeers,gegelia}.
A number of two-nucleon observables 
have been studied in the low-energy theory where pions have been 
integrated out \cite{chen}.
Furthermore, pions can be included explicitly
and treated perturbatively in this scheme \cite{3musketeers}. 
This approach has 
successfully been applied to $NN$ scattering and deuteron
physics \cite{martin}.

The three-nucleon system is a natural test-ground for the understanding
of nuclear forces that has been reached in the two-nucleon system.
Furthermore, it shows some remarkable universal features.
In the context of potential models,
it has been found that different
models of the two-nucleon interaction that are fitted to the same
low-energy two-nucleon ($NN$) data predict different but correlated
values of the triton binding energy $B_3$ and the
$S$-wave nucleon-deuteron ($Nd$) scattering length $a_3^{(1/2)}$ 
in the spin $S=1/2$ channel;
all models fall on a line in the $B_3 \times a_3^{(1/2)}$ plane,
the Phillips line \cite{phillips}.
Other universal features of three-body systems are
the existence of a logarithmic spectrum of bound states
that accumulates at zero energy as the 
two-particle scattering length $a_2$ increases
(the Efimov effect \cite{efimov}), and the collapse of 
the deepest bound state 
when the range of the two-body interaction $r_2$ goes to zero 
(the Thomas effect \cite{thomas}).
These universal features have no obvious explanation in
the framework of conventional potential models.
(Alternative explanations of these features 
that are likely equivalent to ours can be found in Ref. \cite{various}.) 

The critical issue in the extension of any theory from the two- to the
three-body sector is the relative size of three-body forces.
A naive dimensional analysis argument that does not incorporate
the fine tuning in the two-body sector suggests that three-body forces 
appear only in higher orders in the $kR$ expansion.
We have shown that this is supported by $Nd$ data in the
$S=3/2$ channel \cite{3stooges}.
However, for a system of three bosons a non-trivial renormalization
enhances the size of three-body forces \cite{bosons}.
(For a different approach, see Ref. \cite{gegelia3}.)
Here we will show that the renormalization of the three-nucleon
system in the $S=1/2$ channel is nonperturbative and requires a single 
three-body force at leading order.
With such a force, the EFT allows to understand the above
mentioned universal features in a unified way.
Being $SU(4)$ symmetric, this force also supports the view of
spin-flavor as an approximate symmetry of nuclei \cite{wigner}. 
Moreover, if its single parameter is fitted to $a_3^{(1/2)}$,
we find a good description of the energy dependence of scattering
and of triton binding.

\section{Preliminaries}
In order to avoid the difficulties due to the long-range Coulomb force,
we concentrate here on the neutron-deuteron ($nd$) system. 
For simplicity, we also 
restrict ourselves to scattering below the deuteron breakup threshold
where $S$-waves are dominant. There is only one low-energy scale,
$\sqrt{MB_2}\approx 40$ MeV, where $B_2$ is the binding energy of the 
deuteron and $M$ the nucleon mass. Since $\sqrt{MB_2}$ is small compared 
to the pion mass, the pions can be integrated out, only nucleons 
remaining in the EFT as explicit degrees of freedom. The leading order 
of this pionless theory 
\cite{1stooge,3musketeers,gegelia} is equivalent to the leading order
in the KSW counting scheme \cite{3musketeers} where pions are included
perturbatively. 
As a consequence, the extension of our results to higher 
energies is well defined. 

There are two $S$-wave channels for neutron-deuteron scattering,
corresponding to total spin $S=3/2$ and $S=1/2$. For scattering 
in the $S=3/2$ channel all spins are aligned and the two-nucleon
interactions are only in the $^3 S_1$ partial wave. The 
two-body interaction
is attractive but the Pauli principle forbids the three nucleons to be at 
the same point in space. As a consequence, this channel is insensitive 
to short-distance physics and  precise predictions are obtained
in a straightforward way \cite{3stooges}. 
There is also no three-body bound state in 
this channel. The $S=1/2$ channel is more complicated.
The two-nucleon interaction can take place either in the $^3 S_1$
or in the $^1 S_0$ partial waves. This leads to an attractive interaction 
which sustains a three-body bound state, the triton. The $S=1/2$ channel
also shows a strong sensitivity to short-distance physics as the Pauli
principle does not apply. 
We will see that 
the generic features of this channel are thus very similar to the system of 
three spinless bosons. 
There is in the latter a strong cutoff dependence even though
all Feynman diagrams are finite, and renormalization 
requires a leading-order three-body force counterterm \cite{bosons}.

Let us start from the assumption that the three-body force is
of natural size. The lowest-order effective Lagrangian is then given by 
\beqa
\label{lagN}
{\cal L}&=&N^\dagger \left(i\partial_0 +\frac{\vec{\nabla}^2}{2M}\right)N
-C_0^t\left(N^T \tau_2 \vec{\sigma} \sigma_2 N\right)^\dagger \cdot\left(
N^T \tau_2 \vec{\sigma}\sigma_2 N\right) \\
& &-C_0^s\left(N^T \sigma_2 \boldtau \tau_2 N\right)^\dagger \cdot \left(
N^T \sigma_2 \boldtau \tau_2 N\right) +\ldots\,,\nonumber
\eeqa
where the dots represent higher-order terms suppressed by derivatives
and more nucleon fields.
$\vec{\sigma}$ $(\boldtau)$ are Pauli matrices operating in
spin (isospin) space, respectively.
The contact terms proportional to $C_0^t$ ($C_0^s$) correspond to 
two-nucleon interactions in the $^3 S_1$ ($^1 S_0$) $NN$ channels.
Their renormalized values are related to the 
corresponding two-body scattering lengths $a_2^t$ and $a_2^s$ by 
$C_0^{s,t}=\pi a_2^{s,t}/2M$.
Since no derivative interactions appear at this order, this
Lagrangian generates only two-nucleon interactions of zero range.
For practical purposes, it is convenient to rewrite this theory 
by introducing \lq\lq dibaryon'' fields
with the quantum numbers of two nucleons \cite{david}.
In our case, we need two dibaryon fields: 
{\it (i)} a field $\vec{T}$ with
spin (isospin) 1 (0) representing two nucleons interacting in the $^3 S_1$
channel (the deuteron) and 
{\it (ii)} a field $\boldS$ with
spin (isospin) 0 (1) representing two nucleons interacting in the $^1 S_0$
channel.
Using a Gaussian path integration, it is straightforward to show
that the Lagrangian (\ref{lagN}) is equivalent to
\beqa
\label{lagd}
{\cal L}&=&N^\dagger \left(i\partial_0 +\frac{\vec{\nabla}^2}{2M}\right)N
+ \Delta_T \vec{T}^\dagger \cdot\vec{T} +\Delta_S \boldS^\dagger \cdot\boldS\\ 
& &-\frac{g_T}{2}\left( \vec{T}^\dagger \cdot N^T \tau_2 \vec{\sigma} 
\sigma_2 N +h.c.\right) -\frac{g_S}{2}\left(\boldS^\dagger \cdot N^T 
\sigma_2 \boldtau \tau_2 N +h.c.\right) +\ldots \nonumber
\eeqa
At first it may look like the Lagrangian (\ref{lagd}) contains more
parameters than the original one, Eq. (\ref{lagN}). However, 
the scales $\Delta_T$ and $\Delta_S$ are arbitrary and included in 
Eq. (\ref{lagd}) only to give the dibaryon fields the usual mass dimension 
of a heavy field. They can easily be removed by rescaling the dibaryon fields.
All observables depend only on the ratios $g^2_{T,S}/\Delta_{T,S}$.
The introduction of these dibaryon fields makes explicit the formal similarity
to the Amado model \cite{amante}.
We stress, however, that the introduction of 
``quasi-particle'' fields $\vec{T}$ and $\boldS$ carries here 
{\it no} dynamical assumptions \cite{1stooge}. 

\begin{figure}[htb]
\begin{center}
\epsfxsize=10cm
\centerline{\epsffile{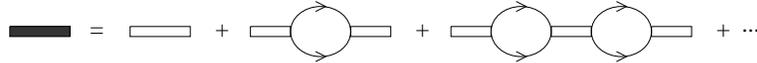}}
\end{center}
\caption{Dressing of the bare dibaryon propagator.}
\label{fig:dress}
\end{figure}

Since the theory is nonrelativistic, all particles propagate forward in time,
the nucleon tadpoles vanish, and the propagator for the nucleon fields is
\beq
\label{nucprop}
iS(p)=\frac{i}{p_0-p^2/2M +i\epsilon}\,.
\eeq
The dibaryon propagators are more complicated because of the coupling
to two-nucleon states. The bare dibaryon propagator is simply a constant, 
$i/\Delta_{S,T}$, but the full propagator gets dressed by nucleon loops
to all orders as illustrated in Fig. \ref{fig:dress}.
The nucleon-loop integral has a linear 
ultraviolet (UV) divergence which can be absorbed 
in $g^2_{T,S}/\Delta_{T,S}$, a finite piece determined by the unitarity 
cut, and subleading terms that have already been omitted in Eq. (\ref{lagd}). 
Summing the resulting geometric series leads to
\beq
\label{Dprop}
i D_{S,T}(p) =  \frac {-i}{- \Delta_{S,T}
             + \frac{M g_{S,T}^{2}}{2\pi} 
               \sqrt{-M p^0+\frac{\vec{p}^{\,2}}{4}-i\epsilon} +i\epsilon}\,,
\eeq
where $g_{T,S}$ and $\Delta_{T,S}$ now denote the renormalized parameters.
The $NN$ scattering amplitude in the respective channel is obtained by
attaching external nucleon lines to the dressed propagator.
In the center-of-mass frame, the $S$-wave amplitude in the 
$^1 S_0$, $^3 S_1$ channels for energy $E=k^2/M$ is
\beq
\label{Tamp}
T^{s,t}(k)=\frac{4\pi}{M}\left( -\frac{2\pi\Delta_{S,T}}{M  g_{S,T}^{2}}
-ik \right)^{-1}\,,
\eeq
and the renormalized parameters $g_{T,S}$ and $\Delta_{T,S}$ can
be determined from 
\beq
\label{renpar}
a_2^{s,t}=\frac{M  g_{S,T}^{2}}{2\pi\Delta_{S,T}}\,.
\eeq
Note that in this zero-range approximation, 
\beq
\label{range}
\sqrt{MB_2}=1/a_2^t, 
\eeq
which holds only approximately, the discrepancy coming from range corrections. 

\section{$S=1/2$ $nd$ Scattering and the Triton}
\begin{figure}[htb]
\begin{center}
\epsfxsize=12cm
\centerline{\epsffile{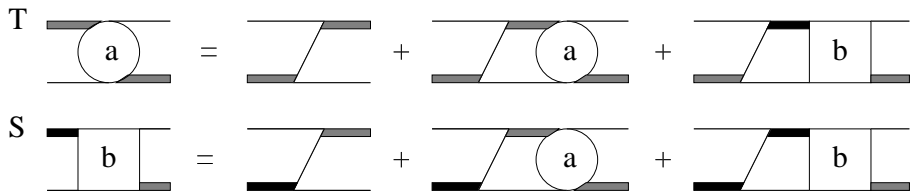}}
\end{center}
\caption{Coupled integral equations for $S=1/2$ $nd$ scattering.
$\vec{T}\,(\boldS)$ dibaryon is indicated by grey shaded (black) thick
line.}
\label{fig:eq12}
\end{figure}

We now use the Lagrangian, Eq. (\ref{lagd}), to describe 
$nd$-scattering in the $S=1/2$ channel and the triton.
The two-nucleon interactions can take place both in the 
$^3S_1$ and $^1S_0$ partial waves. There are two coupled amplitudes, 
$a$ and $b$, as the triton can be built by adding a neutron to
a proton and a neutron which are in either of the two partial waves.
The amplitude $a$ which has both an incoming and outgoing dibaryon
field $\vec{T}$ gives the phase shifts for $S=1/2$ $nd$ scattering;
 $a$ is coupled to the amplitude $b$ which has an
incoming dibaryon $\boldS$ and an outgoing dibaryon $\vec{T}$. 
Although only $a$ corresponds to elastic $S=1/2$ $nd$ scattering,
both amplitudes have the quantum numbers of the triton.
The diagrams for the two amplitudes are obtained 
from the Lagrangian (\ref{lagd}). However,
the leading piece of all diagrams for $a\,(b)$ is of order 
$M g^2_T/Q^2\,(M g_T g_S/Q^2)$, where $Q$ stands for either the external momentum $k$ or
$1/a_2^{s,t}$.
Therefore the diagrams have to be summed to all orders, which 
is conveniently done by solving the coupled integral equations 
shown in Fig. \ref{fig:eq12}.

After the integration over the time component of the loop-momentum
and the projection onto the $S$-waves has been carried out, we have
\begin{eqnarray}
\label{aeq}
& &\frac{3}{2}\left(1/a_2^t+\sqrt{3 p^2 /4 -ME}\right)^{-1}a(p,k)\\
& &=K(p,k)+\frac{2}{\pi}\int_0^\Lambda\frac{q^2 \;dq}{q^2-k^2-i\epsilon}
K(p,q) [a(q,k)+3 b(q,k)] \nonumber\\
\label{beq}
& & 2\,\frac{\sqrt{3 p^2 /4 -ME}-1/a_2^s}{p^2-k^2}\;b(p,k)\\
& &=3 K(p,k)+\frac{2}{\pi}\int_0^\Lambda\frac{q^2 \;dq}{q^2-k^2-i\epsilon}
K(p,q) [3 a(q,k)+b(q,k)]\,.\nonumber
\end{eqnarray}
Here $k$ ($p$) denote the incoming (outgoing) momenta in the center-of-mass 
frame, $M E = 3k^2/4 - (1/a_2^t)^2$ is the total energy, 
and the kernel $K(p,q)$ is given by
\beq
K(p,q)=\frac{1}{2pq}\ln\left(\frac{q^2+pq-p^2-ME}{q^2-pq-p^2-ME}\right)\,,
\eeq
The amplitude $a(p,k)$ is normalized such that $a(k,k)=
(k\cot\delta -ik)^{-1}$ with $\delta$ the elastic scattering phase shift. 
Furthermore, we
have introduced a momentum cutoff $\Lambda$ in the integral equations.
Eqs. (\ref{aeq}, \ref{beq}) have previously been derived 
using a different method \cite{skorny}. In the 
limit $\Lambda\to \infty$ these equations do not have a unique solution
because the phase of the asymptotic solution is undetermined \cite{danilov}.
This is the same feature exhibited by the three-boson systems \cite{bosons}.
For a finite $\Lambda$ this phase is fixed and the solution is
unique. However, the equations with a cutoff display
a strong cutoff dependence that does 
not appear in any order in perturbation theory. The 
amplitude $a(p,k=\mbox{const.})$ shows a strongly oscillating behavior. 
Varying the cutoff $\Lambda$ slightly changes the asymptotic phase 
by a number of ${\cal O}(1)$ and results in large changes of the 
amplitude at the on-shell point $a(p=k)$. This cutoff dependence 
is not created by divergent Feynman diagrams. It is a nonperturbative 
effect and appears although all individual diagrams are 
superficially UV finite.

\begin{figure}[htb]
\begin{center}
\epsfxsize=10cm
\centerline{\epsffile{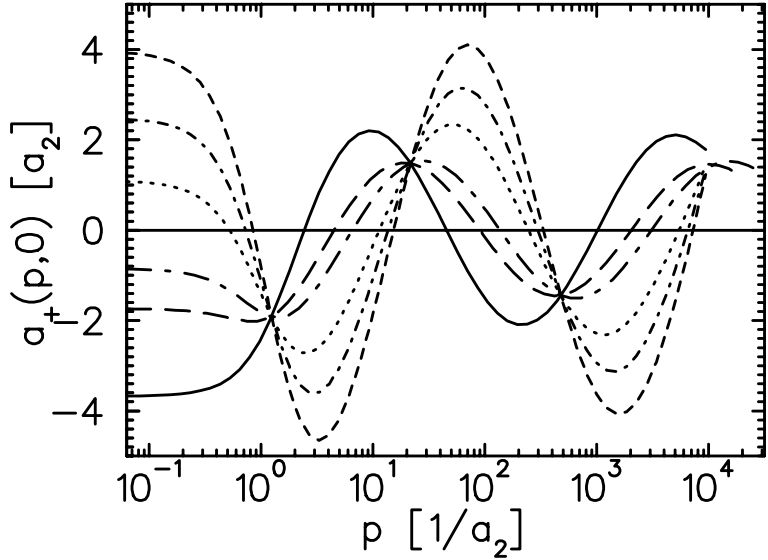}}
\end{center}
\caption{Cutoff dependence of $a_+(p,k=0)$. 
Solid, dashed, and dash-dotted curves are
for $H=0$ and $\Lambda=1.0,\,2.0,\,3.0\times 10^4 a_2^{-1}$, respectively.
Dotted, short-dash-dotted, and short-dashed curves show the effect of the 
three-body force for $\Lambda=10^4 a_2^{-1}$ and $H=-6.0,\,-2.5,\,-1.8$, 
respectively.}
\label{fig:aplus}
\end{figure}

In order to understand this strong $\Lambda$ dependence, 
it is useful to note that the integral equations (\ref{aeq},
\ref{beq}) are $SU(4)$ symmetric in the UV. 
Therefore it is sufficient to consider the $SU(4)$ limit ($a_2^t=a_2^s=a_2$),
since the cutoff dependence is a problem rooted in the 
UV behavior of the amplitudes. Furthermore, we note
that the equations for $a_+=[a+b]$ and $a_-=[a-b]$ decouple in the $SU(4)$ 
limit. The equations for $a_+$ and $a_-$ are
\beqa
\label{apeq}
& &\frac{3}{4}\left(1/a_2+\sqrt{3 p^2/4-ME}\right)^{-1}a_+(p,k) \\
& &= 2K(p,k)
+\frac{2}{\pi}\int_0^\Lambda \frac{q^2\;dq}{q^2-k^2-i\epsilon}
   2K(p,q) a_+(q,k)\nonumber\\
\label{ameq}
& &\frac{3}{4}\left(1/a_2+\sqrt{3 p^2/4-ME}\right)^{-1}a_-(p,k)\\
& &=-K(p,k)\vphantom{\frac{1}{2}}
-\frac{2}{\pi}\int_0^\Lambda \frac{q^2\;dq}{q^2-k^2-i\epsilon}
   K(p,q)\, a_-(q,k)\nonumber\,.
\eeqa
The equation
for $a_+$ is exactly the same equation as in the case of spinless
bosons while the equation for $a_-$ is the same equation as in the 
$S=3/2$ channel. The $S=3/2$ equation is well behaved and its
solution is very insensitive to the cutoff \cite{3stooges}.
Consequently, the cutoff dependence in Eqs. (\ref{aeq}, \ref{beq})
stems solely from the 
equation for $a_+$. As an example, the cutoff dependence of $a_+(p,k=0)$ 
is shown by the solid, dashed, and dash-dotted curves
in Fig. \ref{fig:aplus} for three different cutoffs, $\Lambda=1.0,\,
2.0,\,3.0\times 10^4 a_2^{-1}$. 

Dependence on the cutoff reflects incorrect renormalization,
and thus intolerable dependence on higher-energy modes.
We need to ensure that only low-energy modes contribute
explicitly, otherwise the power counting used to order
two-body interactions ---and in particular to produce
Eq. (\ref{Tamp}) in lowest order regardless of the regularization procedure---
would not hold. 
Now, since the equation for $a_+$ is the same as in the boson case,
we already know the solution to this problem: a one-parameter 
three-body force counterterm $H(\Lambda)/\Lambda^2$ 
that runs with the cutoff $\Lambda$ \cite{bosons}. 
We simply replace
\beq
K(p,k) \rightarrow K(p,k)+\frac{H(\Lambda)}{\Lambda^2}
\eeq
in Eq. (\ref{apeq}). 
The dotted, short-dash-dotted, 
and short-dashed curves in Fig. \ref{fig:aplus} show the effect of the 
three-body force $H(\Lambda)$ on $a_+(p,k=0)$ for $\Lambda=10^4 a_2^{-1}$ 
and $H=-6.0,\,-2.5,\,-1.8$. It is clearly seen that the variation of 
$H$ for a constant $\Lambda$ has the same effect on the amplitude as
varying the cutoff. Consequently, we can compensate the 
changes in the asymptotic phase when $\Lambda$ is varied by adjusting 
the three-body force term appropriately. (A more rigorous discussion of 
the renormalization procedure can be found in Ref. \cite{bosons}.)

\begin{figure}[htb]
\begin{center}
\epsfxsize=10cm
\centerline{\epsffile{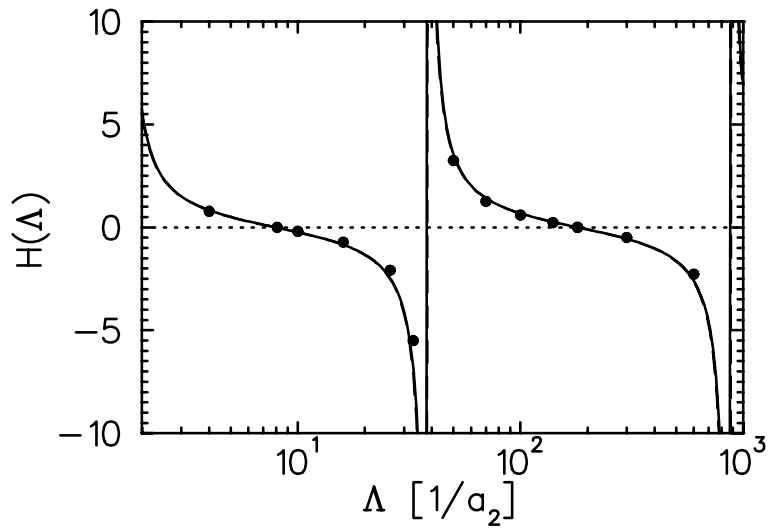}}
\end{center}
\caption{Running of $H(\Lambda)$ for $\Lambda_* = 0.9 \mbox{ fm}^{-1}$:
(a) from Eq. (\ref{runH}) (solid line), (b) from numerical solution of
Eq. (\ref{apeq}) (dots).}
\label{fig:betaf12}
\end{figure}

We can obtain an approximate expression for the running of $H(\Lambda)$ 
from invariance under the renormalization group. Requiring that the 
equation for $a_+$ does not change its form when the high momentum modes 
are integrated out, we find
\beq
\label{runH}
H(\Lambda)=-    \frac{\sin(s_0\ln({\Lambda}/{\Lambda_\star})-
                   {\rm arctg}(1/s_0))}
                 {\sin(s_0 \ln({\Lambda}/{\Lambda_\star})+
                   {\rm arctg}(1/s_0))}
\eeq
where $s_0\approx 1.0064$ 
and $\Lambda_*$ is a dimensionful parameter that determines 
the asymptotic phase of the off-shell amplitude
\cite{bosons}.
The running of the three-body force $H(\Lambda)$ according to
Eq. (\ref{runH}) is shown by the solid line in Fig. \ref{fig:betaf12}.
The dots are obtained by adjusting $H(\Lambda)$ such that the 
low-energy solution of Eq. (\ref{apeq}) remains unchanged when $\Lambda$
is varied. The observed agreement provides a numerical justification
for our procedure. The three-body force
is periodic with $H(\Lambda_n)=H(\Lambda)$ for
$\Lambda_n=\Lambda\exp(n\pi/s_0)\approx \Lambda (22.7)^n$.
In particular, the bare three-body force vanishes
for a discrete set of cutoffs. Note, however, that
invariance under continuous changes
in the cutoff does require a non-vanishing bare three-body force.

An important point should be stressed here. We have first renormalized the
two-body subamplitude (the dibaryon propagator) and then inserted the 
result in the three-body equation. The loop appearing in this equation
was then regulated by a cutoff and renormalized by the introduction of a 
three-body force. This is equivalent to using {\it separate} cutoffs 
$\Lambda$ (for the three-body equation) and $\Lambda'$ (for the dibaryon 
propagator) and taking $\Lambda'\rightarrow\infty$ first (changing the 
two-body parameters $g_{T,S}^2 /\Delta_{T,S}$ as a function of $\Lambda'$ 
to keep the two-body observables
fixed) and then performing the renormalization of the three-body equation 
discussed above. A more standard approach would be to use the same cutoff
for the loops appearing in both the dibaryon propagator and the three-body 
equation. The fact that the approach used in this paper is legitimate  
becomes evident by noting that the details of the renormalization are 
changed from one approach to the other, but not the physical results.
To see that, let us consider the changes due to the use of
$\Lambda=\Lambda'$ throughout.
The two-body subamplitude would then be changed 
by terms suppressed by powers of $Q/\Lambda$, where $Q$ represents the 
typical momentum flowing through the two-body subamplitude. When inserted 
in the three-body equation, these terms are important only for loop momenta 
($q$ in Eqs. (\ref{aeq}, \ref{beq})) of the order of $\Lambda$. 
In this region of integration
other arbitrary choices like, {\it e.g.}, the precise 
shape of the cutoff used (sharp or smooth) are also important. 
All these differences occurring at the cutoff scale are absorbed in the 
three-body force, which would result in a different $\Lambda$ dependence 
than given in  Eq. (\ref{runH}). The physical results for small momenta  
($p\ll\Lambda$), however, are unchanged up to terms suppressed by 
powers of $p/\Lambda$. Those terms depend on our particular choice 
of regulators and also break symmetries like Galilean invariance. 
There are many other terms of the same order in $p/\Lambda$ that were 
already discarded, since we have performed an expansion in powers of $p$.
The error introduced by this omission, however, is of higher order than we 
are working at.

Formally, the three-body force term is obtained by adding 
\beqa
\label{3bod}
{\cal L}_3 = -\frac{2MH(\Lambda)}{\Lambda^2} \! \! \! \! \! \!
  && \bigg( g_T^2 N^\dagger 
(\vec{T}\cdot\vec{\sigma})^\dagger (\vec{T}\cdot\vec{\sigma}) N \\
& & \; +\frac{1}{3} g_T g_S \left[ N^\dagger (\vec{T}\cdot\vec{\sigma})^\dagger
(\boldS\cdot\boldtau) N + h.c. \right] \nonumber \\
& & \; + g_S^2 N^\dagger (\boldS\cdot\boldtau)^\dagger (\boldS\cdot
\boldtau) N \bigg)\,,\nonumber
\eeqa
to the Lagrangian (\ref{lagd}).
Eq. (\ref{3bod}) represents a contact three-body force written in
terms of dibaryon and nucleon fields. Via a Gaussian
path integration it is equivalent to a true three-nucleon force,
\beqa
\label{3bodN}
{\cal L}_3 = &-&\frac{2MH(\Lambda)}{\Lambda^2}\bigg(\,
\frac{g_T^4}{4\Delta_T^2}\,(N^T \tau_2 \sigma_k \sigma_2 N)^\dagger
(N^\dagger \sigma_k \sigma_l N)(N^T \tau_2 \sigma_l \sigma_2 N)\\
& &+\frac{1}{3}\frac{g_T^2}{2\Delta_T}\frac{g_S^2}{2\Delta_S}
\left[ (N^T \tau_2 \sigma_k \sigma_2 N)^\dagger
(N^\dagger \sigma_k \tau_l N)(N^T \sigma_2 \tau_l \tau_2 N) +h.c. \right]
\nonumber\\
& &+\frac{g_S^4}{4\Delta_S^2}\,(N^T \sigma_2 \tau_k \tau_2 N)^\dagger
(N^\dagger \tau_k \tau_l N)(N^T \sigma_2 \tau_l \tau_2 N)\,\bigg)\,.
\nonumber
\eeqa
By performing a Fierz rearrangement, it can then be shown that the three 
terms in Eq. (\ref{3bodN}) are equivalent. 
As a consequence, there is
only  one three-body force which is also $SU(4)$ symmetric.
Therefore the choice of the three-body force, Eq. (\ref{3bod}),
is by no means arbitrary. In fact, it is the only $S$-wave three-body
force with no derivatives that can be written down.
This can be seen by remembering that three fermions (antifermions)
in a completely symmetric spatial wave need to be in a completely 
antisymmetric spin-isospin state which is a $\bar{\bf 4}\,({\bf 4})$ 
of $SU(4)$. Consequently, the three-body force transforms as
${\bf 4}\otimes\bar{\bf 4}={\bf 1}\oplus{\bf 15}$, but only the 
singlet is separately invariant under the spin and isospin subgroups.
The two-body parameters $g_T^2/\Delta_T$
and $g_S^2/\Delta_S$ appear in Eq. (\ref{3bod}) because the effect of
the three-body force cannot be separated from the effect of the 
two-body interaction. Both are linked by the renormalization procedure.
Naive power counting would suggest that the three-nucleon force 
scales with $1/(M m_\pi^4)$. The three-body force from Eqs. 
(\ref{3bod}, \ref{3bodN}), however, is enhanced by the renormalization 
group flow by two powers of $a_2$. Using Eq. (\ref{renpar}), it is found to  
scale as $a_2^2/(M m_\pi^2)$ which makes it leading order.
$H(\Lambda)$ contains one new dimensionful parameter, $\Lambda_*$,
which must be calculated from QCD or determined from experiment.

Recently Mehen, Stewart, and Wise \cite{caltech}
remarked that the two-nucleon 
amplitude (\ref{Tamp}) is approximately $SU(4)$ symmetric
for $p\gg 1/a_2$. They also noticed that the only
$S$-wave four-nucleon force that can be written down
is $SU(4)$ symmetric.
Furthermore, there are no contact interactions with more than four 
nucleons without derivatives because of the Pauli principle.
It is also reasonable to assume that the low-energy dynamics of 
other nuclei
is dominated by $S$-wave interactions,
as is the case for the deuteron and the triton. Therefore our 
$SU(4)$ symmetric three-body force together with the findings of
Mehen {\it et al.} \cite{caltech} gives an explanation 
for the approximate $SU(4)$ symmetry \cite{wigner}
in nuclei.

Now we are in position to solve the full equations for the 
broken $SU(4)$ case. Including  the three-body force from above
into Eqs. (\ref{aeq}, \ref{beq}), we have
\begin{eqnarray}
\label{aeqf}
& &\frac{3}{2}\left(1/a_2^t+\sqrt{3 p^2 /4 -ME}\right)^{-1}a(p,k)=
K(p,k)+\frac{2H(\Lambda)}{\Lambda^2}\\
& &+\frac{2}{\pi}\int_0^\Lambda\frac{q^2 \;dq}{q^2-k^2-i\epsilon}
\left[K(p,q) [a(q,k)+3 b(q,k)] +\frac{2H(\Lambda)}{\Lambda^2}[a(q,k)+b(q,k)]
\right]\nonumber\\
\label{beqf}
& & 2\,\frac{\sqrt{3 p^2 /4 -ME}-1/a_2^s}{p^2-k^2}\;b(p,k)=
3 K(p,k)+\frac{2H(\Lambda)}{\Lambda^2}\\
& &+\frac{2}{\pi}\int_0^\Lambda\frac{q^2 \;dq}{q^2-k^2-i\epsilon}
\left[K(p,q) [3 a(q,k)+b(q,k)] +\frac{2H(\Lambda)}{\Lambda^2}[a(q,k)+
b(q,k)]\right]\,.
\nonumber
\end{eqnarray}
We need one three-body datum to fix the three-body force parameter 
$\Lambda_*$. We choose the experimental value for the $S=1/2$ $nd$ scattering 
length, $a_3^{(1/2)}=(0.65 \pm 0.04)\mbox{ fm}$ \cite{dilg},
and find $\Lambda_* = 0.9 \mbox{ fm}^{-1}$. (For the special cutoffs with 
vanishing $H(\Lambda)$, we recover the results of Ref. \cite{kharchenko}.)
Although one three-body datum is needed as input,
the EFT has not lost its predictive power. We can still predict 
{\it (i)} the
energy dependence of $S=1/2$ $nd$ scattering and 
{\it (ii)} the binding
energy of the triton. 

\begin{figure}[htb]
\begin{center}
\epsfxsize=10cm
\centerline{\epsffile{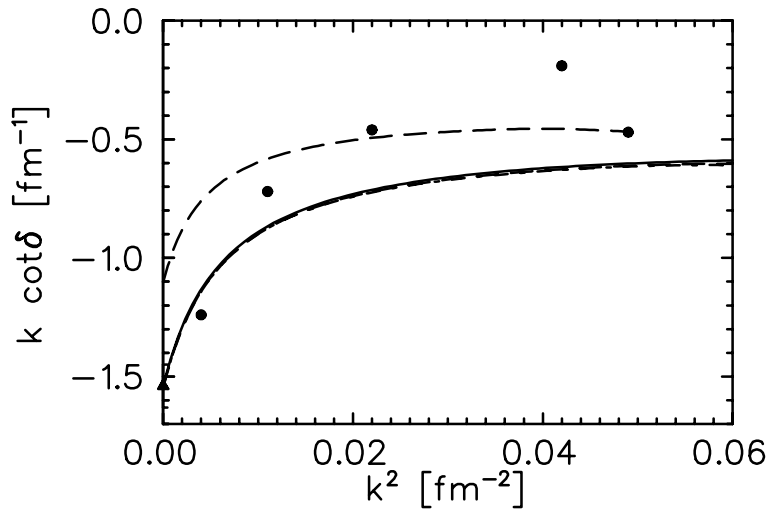}}
\end{center}
\caption{Energy dependence for $S=1/2$ $nd$ scattering with
three different cutoffs $\Lambda=1.9,\, 6.0,\, 11.6 \mbox{ fm}^{-1}$
(hidden under the solid curve)
for $\Lambda_*=0.9\mbox{ fm}^{-1}$.
Dashed curve gives estimate of range corrections. 
Data are from the phase-shift analysis 
of van Oers and Seagrave \protect\cite{vOers} (dots) and 
a measurement by Dilg {\it et al.} \protect\cite{dilg} (triangle).}
\label{fig:ed12}
\end{figure}

The resulting energy dependence of $S=1/2$ $nd$ scattering
for three different cutoffs is shown in Fig. \ref{fig:ed12}.
It is clearly seen that the introduction of the three-body force 
renders the low-energy amplitude cutoff independent.
The scattering length is reproduced exactly because it was used
to fix $\Lambda_*$. The agreement for finite momentum is at least
encouraging. Our experience from the $S=3/2$ channel is that
the range corrections improve the agreement considerably 
\cite{3stooges}. The dashed curve in Fig. \ref{fig:ed12} gives 
a crude estimate of these corrections. 
In our calculations we
take the deuteron binding energy $B_2$ from experiment 
and determine $a_2^t$ from Eq. (\ref{range}). The dashed curve in Fig.
\ref{fig:ed12} is obtained by taking the experimental value of $a_2^t$
as input and leaving all other parameters unchanged.
{}From the estimated size of the range corrections, we anticipate an
improved agreement once these corrections are included. (One should also
keep in mind that the experimental phase-shift analysis \cite{vOers}
does not give any error estimate.)

\begin{figure}[htb]
\begin{center}
\epsfxsize=10cm
\centerline{\epsffile{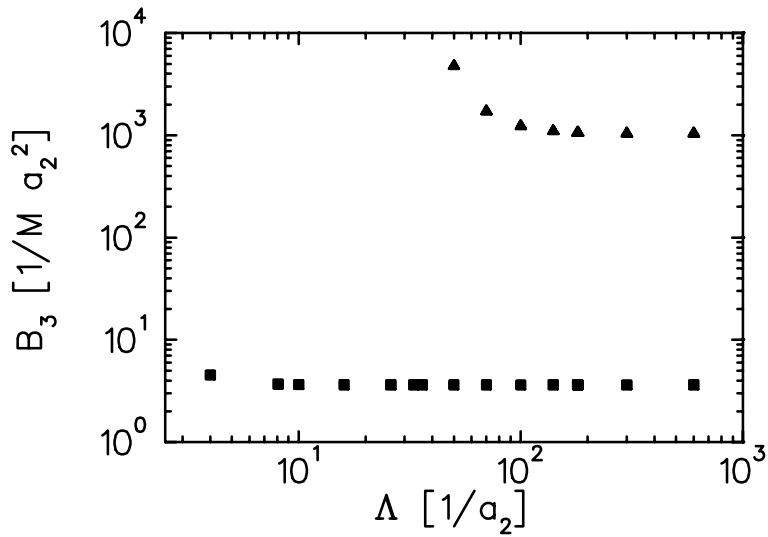}}
\end{center}
\caption{Three-nucleon bound state spectrum for $\Lambda_*=0.9\mbox{ fm}^{-1}$.
The shallowest bound state corresponds to the triton.}
\label{fig:bs12}
\end{figure}

The triton binding energy is obtained
from the solution of the homogeneous equations corresponding to
Eqs. (\ref{aeqf}, \ref{beqf}) for $E=-B_3$.
In Fig. \ref{fig:bs12} we show the bound state spectrum as a function of 
the cutoff $\Lambda$ for a particular $\Lambda_*$. 
The shallowest bound state is the triton. Its binding energy is cutoff
independent  as long as $\Lambda \gg 1/a_2$. However, as $\Lambda$ is increased
new deeper bound states appear whenever $H(\Lambda)$
goes through a pole. 
These new bound states appear with infinite
binding energy directly at the pole. As the cutoff is further increased,
their binding energy decreases and becomes cutoff independent as well. 
The poles of $H(\Lambda)$ can be 
parametrized as
\beq
\label{poles}
\Lambda_n=\underbrace{f(\Lambda_* a_2)}_{{\cal O}(1)}
\exp(n\pi/s_0) a_2^{-1}\,.
\eeq
One counts $n$ bound states for $\Lambda_{n-1} 
< \Lambda < \Lambda_n$. However, only the states between threshold and 
$\Lambda\sim m_\pi \sim 1/R$ are within the range of the EFT. By solving
Eq. (\ref{poles}) for the number of bound states $n$, we then obtain
\beq
\label{efimovspec}
\#_{BS}=\frac{s_0}{\pi}\ln\left(\frac{a_2}{R}\right)
+ {\cal O}(1)\,,
\eeq
and recover the well-known Efimov effect \cite{efimov}.
In the limit $a_2 \to \infty$, an infinite number of 
shallow three-body bound states accumulates at threshold.
That these bound states are shallow follows from the fact that 
the binding energy is naturally given in units of 
$1/(M a_2^2)$ which vanishes as $a_2 \to \infty$.
Furthermore, we also recover the Thomas effect \cite{thomas}.
In a hypothetical world where
$\Lambda \sim 1/R \to \infty$, the range of the EFT
increases and deeper and deeper physical bound states appear. 
Consequently, there is an infinitely deep bound
state for $\Lambda \to \infty$. However, for $m_\pi \sim 1/R$ as
in the real world the deep bound states are outside the range of the EFT
and their presence does not influence the physics of the shallow ones. 

\begin{figure}[htb]
\begin{center}
\epsfxsize=10cm
\centerline{\epsffile{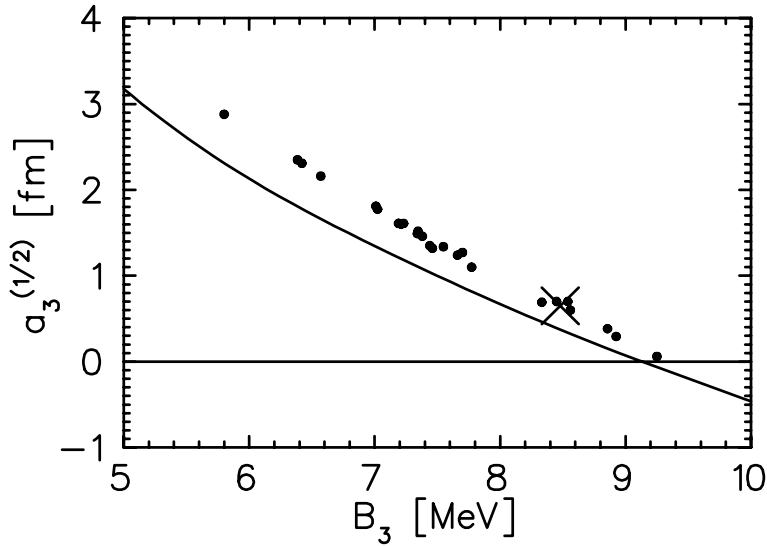}}
\end{center}
\caption{$S=1/2$ $nd$ scattering length $a_3^{(1/2)}$
as function of the triton binding energy $B_3$ (Phillips line): 
EFT to leading order 
(solid line); potential models (dots); and experiment (cross).}
\label{fig:phill}
\end{figure}

Moreover, the variation of the parameter $\Lambda_*$ gives a natural 
explanation for the Phillips line \cite{phillips}.
The three-nucleon system has been studied with
many phenomenological two-nucleon potentials where
particles can have momenta as high as 1 GeV.
Different phase-equivalent two-body potential models have 
different off-shell amplitudes,
which are equivalent to different three-body forces.
In some cases explicit three-body potentials have also been added.
These models can be viewed as particular examples of high-energy dynamics.
At low energies, they must all be equivalent to the EFT 
with particular values for its parameters.
What we have shown is that in leading order they can only differ 
in the value of the {\it single} three-body parameter $\Lambda_*$.
Varying $\Lambda_*$ then generates a {\it line}.
In Fig. \ref{fig:phill} we show the Phillips line obtained in the 
EFT compared with results from various potential model calculations
\cite{jim} and the experimental values for $B_3$  and $a_3^{(1/2)}$.
Our Phillips line is slightly below the one from the 
potential models. We expect this discrepancy to be reduced
once range corrections are taken into account.

The dynamics of QCD chooses a particular value 
of $\Lambda_*$, which up to higher order
corrections is $\Lambda_*=0.9\mbox{ fm}^{-1}$.
For this value
we find $B_{3} = 8.0$ MeV for the triton binding energy, 
to be compared with the experimental result
$B_3^{exp} = 8.48$ MeV, which is known to very high precision. 
The theory without pions seems to work for triton physics. However,
to draw definite conclusions one has to include the range corrections
for both the energy dependence and the triton binding energy.

The three-body force (\ref{3bodN})
involves three nucleons at the same point 
(within the resolution of the EFT) 
in a relative $S$ wave. 
It cannot contribute to the $S=3/2$ channel where
the three nucleons spin in the same direction.
Indeed, we have found that the quartet can be well predicted
with two-nucleon input alone \cite{3stooges}. 
We stress that the three-body force that arises at 
low energies is not the same that appears at higher energies.
As momentum increases, part of the three-body contact operator (\ref{3bodN})
may be resolved into iterations of the two-body potential.
The strong renormalization of this operator at low energies
is not necessarily in contradiction with 
the potential-model-based phenomenology where 
three-body forces are found to be small.

\section{Conclusions}
We have studied the three-nucleon system using EFT methods. While for
$nd$ scattering in the $S=3/2$ channel precise predictions are obtained 
in a straightforward way \cite{3stooges}, 
the $S=1/2$ channel is more complicated.
It displays a strong cutoff dependence even though all individual
diagrams are UV finite.
In this channel a nonperturbative renormalization takes place 
similar to the case of spinless bosons. This renormalization
group argument requires an $SU(4)$ symmetric three-body force which is
enhanced to leading order. Together with 
the recent observations of Mehen {\it et al.} \cite{caltech}
this gives an explanation for the approximate $SU(4)$ symmetry in nuclei.
Furthermore, we find that the Phillips line is a consequence of 
variations in the new dimensionful parameter $\Lambda_*$ which is 
introduced by the three-body force. $\Lambda_*$ is not determined by
low-energy two-nucleon data alone and has to be determined from a 
three-body datum. 

This is to be contrasted to other approaches to the problem that
model the two-body interaction at distances of the order of the
effective range as, for instance, the Amado model \cite{amante}. 
There a form factor is introduced in the interaction between two nucleons 
whose parameters are chosen to reproduce not only the two-body scattering 
length but also the effective range. Since the effective range expansion is 
not to be trusted at momenta of the order of the inverse effective range,
this corresponds, in our language, to taking the three-body force
equal to zero and picking a particular form for the regulator (the 
effective range term in the two-body subamplitude acts as a cutoff). 

The leading order of the EFT gives a quantitative description of
the triton binding energy and the energy dependence for $S=1/2$
$nd$ scattering. 
In the appropriate limits for the two-body 
parameters $a_2$ and $R$,
we also recover the well known Thomas and Efimov effects.
The theory shows the potential for a realistic 
description of the triton once range corrections are included.
Immediate applications of the EFT include polarization observables 
in $nd$ scattering and triton properties such as its charge form
factor. The incorporation of the long-range Coulomb force would widen the 
possible applications considerably as $pd$ scattering and the
physics of $^3$He becomes accessible. Work on these extensions is
in progress.

Finally, since in leading order the pionless theory is equivalent 
to a theory with explicit (perturbative) pions \cite{3musketeers},
the success in the three-nucleon system opens the possibility
of applying the EFT method to a large class of systems with three
or more nucleons.

\section*{Acknowledgements}
We would like to thank Jim Friar for the potential model data.
This research was supported in part by the U.S. Department of Energy
grants DOE-ER-40561 and DE-FG03-97ER41014, the Natural Sciences and 
Engineering Research Council of Canada, and the U.S. National Science
Foundation grant PHY94-20470.

\end{document}